# Construction graphique d'entrepôts et de magasins de données

**Frédéric Bret, Olivier Teste**

IRIT (Institut de Recherche en Informatique de Toulouse),
équipe SIG, Université Toulouse III,
118, Route de Narbonne - 31062 Toulouse cedex 04, France

Email : {bret, teste}@irit.fr

**Résumé :**

De nos jours, les systèmes décisionnels sont devenus un thème de recherche important dans le domaine des bases de données. Les entrepôts de données ("*data warehouses*") et les magasins de données ("*data marts*") constituent les éléments principaux de ces systèmes.

Cet article présente notre architecture de systèmes d'aide à la décision. Nous proposons des interfaces graphiques qui aident l'administrateur du système décisionnel lors du processus de construction de l'entrepôt et des magasins de données. Nous présentons une interface de construction de l'entrepôt de données basée sur un modèle conceptuel orienté objet. Celui-ci permet l'historisation des données de l'entrepôt à trois niveaux : attribut, classe et environnement. Nous présentons également une interface de construction des magasins de données permettant de réorganiser les données de l'entrepôt selon un modèle orienté objet multidimensionnel.

**Mots clés :**

entrepôts de données, magasins de données, historisation, multidimensionnel, interfaces graphiques pour l'administrateur.

**Abstract :**

Nowadays, decisional systems have became a significant research topic in databases. Data warehouses and data marts are the main elements of such systems.

This paper presents our decisional support system. We present graphical interfaces which help the administrator to build data warehouses and data marts. We present a data warehouse building interface based on an object-oriented conceptual model. This model allows the warehouse data historisation at three levels: attribute, class and environment. Also, we present a data mart building interface which allows warehouse data to be reorganised through a multidimensional object-oriented model.

# 1 Introduction

Issus de l'industrie pour servir de base à l'analyse des données, les entrepôts de données ("*data warehouses*") sont devenus un thème de recherche à part entière [*WIDO95*] [*CHAU97*] [*WU97*]. L'objectif de ces entrepôts est de servir de support à la prise de décision en permettant une meilleure exploitation des informations contenues dans les systèmes opérationnels des entreprises.

*Un entrepôt est une collection de données intégrées, orientées sujet, non volatiles, historisées, résumées et disponibles pour l'interrogation et l'analyse* [*INMO96*]. L'entrepôt intègre les données pertinentes - nécessaires à la prise de décision - des systèmes opérationnels de l'entreprise. Cette intégration fait appel à des processus d'unification de données hétérogènes, d'agrégation et d'historisation.

L'entrepôt de données est généralement destiné aux décideurs de l'entreprise souvent non spécialistes de l'informatique. L'entrepôt doit être muni d'interfaces graphiques adaptées à des utilisateurs non informaticiens

Le contexte applicatif de nos travaux concerne le milieu médical (sécurité sociale). En effet, dans le cadre de l'Assurance Maladie en France, les systèmes opérationnels contiennent de très importants volumes de données relatifs aux assurés sociaux, aux professionnels de santé, aux établissements médicaux. Ces données sont gérées par les Centres de Traitements Informatiques[*]. L'exploitation de ces données par les experts de l'Assurance Maladie, tels que les médecins conseils, s'effectue de manière parfois empirique par des moyens classiques (vues, requêtes SQL, outils graphiques d'interrogation,…). Ce processus d'exploitation s'avère souvent fastidieux et n'apporte pas toujours des réponses satisfaisantes. Les entrepôts de données peuvent apporter des solutions pour une meilleure exploitation de ces données.

L'objectif de cet article est de présenter notre architecture des systèmes décisionnels en distinguant les problématiques liées à la construction de tels systèmes. En particulier, nous présentons deux interfaces graphiques dédiées à la construction d'entrepôts et de magasins de données.

Cet article est organisé en quatre sections. Dans la section 2, nous présentons des travaux de recherche dans le domaine des entrepôts de données et dans le domaine des langages graphiques destinés aux bases de données orientées objet. Dans la section 3, nous présentons notre architecture de systèmes décisionnels à quatre niveaux. Dans la section 4, nous proposons un outil graphique permettant la construction d'entrepôts de données (correspondant à la seconde phase de notre processus). Cet outil s'appuie sur un modèle conceptuel orienté objet permettant la gestion de données "*historisées*". La section 5 propose un outil graphique permettant la construction de magasins de données (correspondant à la troisième phase de notre processus). Cet outil est basé sur un modèle conceptuel orienté objet *multidimensionnel* et sur une *démarche* de construction.

## 2 Travaux existants

Les entrepôts de données sont issus originellement du monde industriel. Depuis quelques années, les travaux de recherche sur ce thème se sont multipliés. Par conséquent, nous présentons dans cette section les principaux outils utilisés dans l'industrie, puis divers projets et travaux de recherche sur les entrepôts.

### 2.1 Outils de l'industrie

Les outils issus de l'industrie se concentrent essentiellement sur les aspects de l'interrogation. Plusieurs requêteurs (Business Objects, Impromptu, Discover2000) et outils OLAP (Powerplay, Discover, Essbase, Express, Mercury) sont proposés. Les requêteurs facilitent la construction de requêtes SQL (jointures masquées, alias de noms d'attributs,…). Les outils OLAP se concentrent sur l'analyse multidimensionnelle en proposant des objets graphiques (tableaux croisés,…). Ces outils utilisent les données de l'entrepôt ou des magasins, mais ne permettent pas leur construction.

Peu d'outils sont disponibles pour assister l'administrateur dans la construction des entrepôts et des magasins de données (Data Mart Builder, SAS/Warehouse Administrator, Warehouse Manager). Avec ces outils, le schéma de l'entrepôt ou du magasin doit être construit au préalable. L'administrateur a la charge d'associer chaque attribut cible de ce schéma à un attribut d'une source de données. Cette phase reste complexe et demande une connaissance importante des structures sources.

Notre approche se distingue en proposant de construire *incrémentalement* l'entrepôt ou le magasin ; l'administrateur construit l'entrepôt en dérivant les données sources utiles. Cette construction se fait à partir d'une représentation graphique des sources. Ainsi, l'administrateur peut appréhender plus facilement la structure et la sémantique des données sources.

### 2.2 Travaux de recherche

Le projet WHIPS [*WIEN96*] propose une architecture des systèmes décisionnels centrée sur l'intégration des sources de données et la maintenance incrémentale de l'entrepôt. L'objectif du projet WHIPS est de proposer des algorithmes permettant la maintenance de vues matérialisées. [*SAMO98*] s'intéresse aux problèmes liés à l'intégration des sources de données en s'inspirant de travaux effectués dans les BD fédérées. Dans [*JARK98*] une architecture DWQ (Data Warehouse Quality) à trois niveaux (conceptuel, logique, physique) est proposée. L'objectif est de définir des critères de qualité pour une meilleure organisation des méta-données de l'entrepôt.

Initialement, le concept de vue [*GARD93*] a été développé dans le domaine des bases de données pour exprimer des contraintes de confidentialité sur les données. Dans les entrepôts de données, les vues sont *matérialisées* [*GUPT95*]. Cette technique consiste à calculer la vue et à stocker le résultat dans l'entrepôt. De nombreux travaux tels que [*ZHUG95*] [*BELL98*] [*YANG98*] proposent des algorithmes relatifs à la maintenance incrémentale et à la réduction des coûts de maintenance des vues matérialisées. Tous ces travaux utilisent une définition textuelle des vues (sous la forme SPJ : Sélection - Projection - Jointure) et n'offrent pas une solution basée sur une interface graphique comme nous le proposons.

D'autres travaux de recherche concernent le processus OLAP. Ce processus consiste à organiser les données pour faciliter et optimiser leur analyse. Lors de l'analyse, les données sont regroupées, résumées, consolidées et visualisées selon différents niveaux de détail. Les travaux relatifs au processus OLAP se situent à un niveau *conceptuel* [*GOLF98*], [*GYSS97*], [*LEHN98*], *logique* [*AGRA97*], [*GRAY96*], [*REDB95*] et *physique* [*COLL96*], [*ZHAO97*].

Avec la généralisation de l'informatique dans tous les domaines, les langages basés sur un paradigme à caractère visuel se développent naturellement dans les différents systèmes informatiques. Puisque nos travaux s'effectuent dans un cadre orienté objet, nous nous limitons aux bases de données orientées objet en distinguant :

- les *langages tabulaires* comme OOQBE [*STAE91*], VQL [*VADA93*],
- les *langages iconiques* comme VISTA [*BELI96*],
- les *langages graphiques* comme OHQL [*ANDO91*], VOHQL [*ANDO97*], EIGOO [*SAYA98*].

Les langages graphiques présentent l'avantage de visualiser le schéma de la base avec les différents liens entre les classes, tandis que les langages tabulaires présentent un apport visuel limité et que les langages iconiques font mal apparaître les différents liens entre les classes.

## 3 Notre architecture

Nous proposons une architecture de systèmes d'aide à la décision. Elle est basée sur une organisation à quatre niveaux (*cf.* figure 1), chacun d'eux étant associé à un modèle de données particulier. Par exemple, le modèle de l'entrepôt gère l'évolution des données (données historisées) tandis que le modèle des magasins organise les données de manière multidimensionnelle pour optimiser l'analyse.

Nous utilisons un modèle orienté objet dont la sémantique plus riche permet une meilleure gestion des données à structures complexes et multimédias.

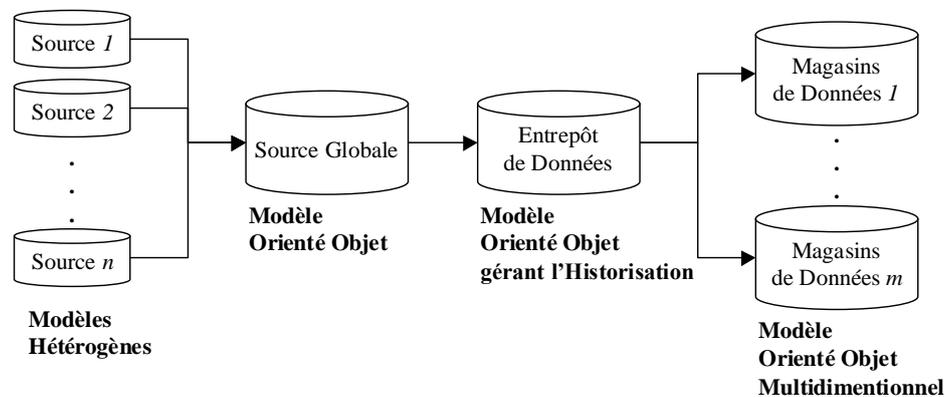

**Figure 1.**   *Notre architecture de systèmes d'aide à la décision.*

Dans cet environnement, nous proposons un processus permettant la construction du système d'aide à la décision. Ce processus se décompose en trois phases permettant d'isoler les différents problèmes à résoudre : une phase d'*intégration* des sources de données, une phase de *construction* de l'entrepôt et une phase de *réorganisation* des données pour construire les magasins.

La phase *d'intégration* permet de générer une source de données globale obtenue à partir de l'intégration des différentes sources hétérogènes et distribuées. Cette phase doit permettre de résoudre les problèmes liés à l'hétérogénéité et à la distribution des sources. Elle fait appel entre autre à des techniques définies pour les bases de données fédérées [*SAMO98*] et les bases de données réparties [*RAVA96*].

La phase de *construction* de l'entrepôt de données utilise le mécanisme des vues matérialisées. Nous utilisons le mécanisme des vues pour permettre à l'administrateur d'indiquer simplement au système quelles sont les données sources nécessaires à la construction de l'entrepôt. Cette phase consiste à définir une vue sur la source de données globale au travers d'une interface graphique qui assiste l'administrateur. Une opération importante de cette phase consiste à déterminer les données dont les évolutions doivent être conservées dans l'entrepôt. Le système génère ensuite l'entrepôt de données comme une base de données orientées objet contenant les données extraites de la source. L'administrateur détermine une période de rafraîchissement pour maintenir l'entrepôt de données consistant avec les sources.

La phase de *réorganisation* consiste à générer à partir de l'entrepôt différents magasins de données basés sur un modèle multidimensionnel. Chaque magasin prépare les données pour l'analyse (OLAP – On Line Analytical Processing) ; les données sont organisées autour d'un centre d'intérêt (le "*fait*") tandis que les dimensions constituent les différents axes de l'analyse. Cette phase aborde des problèmes liés à la structure des données stockées dans les magasins. Cette structure permet l'analyse optimale d'informations orientées "*sujet*".

Différentes interfaces graphiques sont mises à la disposition de l'administrateur à chaque phase du processus de création du système d'aide à la décision. Les outils graphiques proposés sont destinés à des utilisateurs experts (nous utiliserons le terme d'administrateur), non nécessairement informaticiens, qui connaissent parfaitement les besoins des utilisateurs finals du système décisionnel. L'administrateur a trois tâches principales : concevoir, construire et maintenir le système décisionnel.

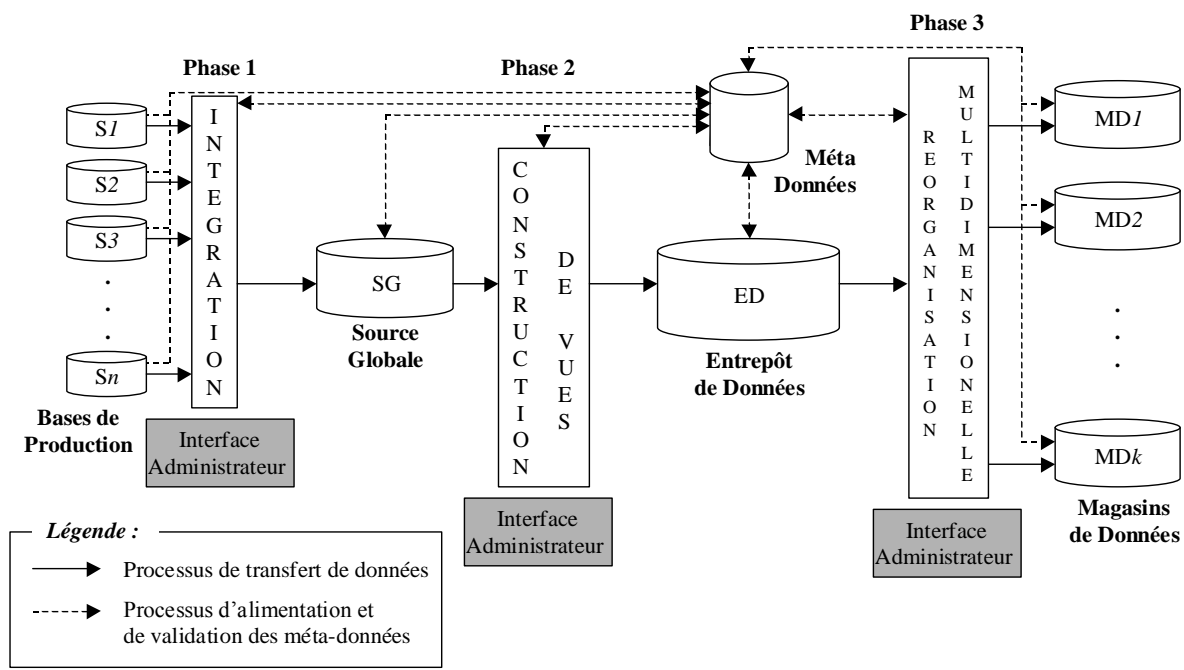

**Figure 2.** *Processus de construction des systèmes d'aide à la décision.*

## 4  Construction graphique de l'entrepôt de données

Cette section décrit la phase de construction de l'entrepôt de données ; ce dernier est basé sur une extension d'un modèle orienté objet qui intègre des concepts relatifs aux entrepôts de données (typologie d'attributs et d'opérations). La gestion de données temporelles se fait à trois niveaux (attribut, classe d'objets et ensemble de classes d'objets). [*TEST99*] décrit en détail notre modèle conceptuel d'entrepôts de données. L'outil permettant à l'administrateur de construire l'entrepôt repose sur la démarche suivante :

- 1 - Choisir un schéma source.
- 2 - Projeter les classes du schéma source dans l'entrepôt de données et Sélectionner les instances sources qui alimentent les classes de l'entrepôt ;
- 3 - Enrichir l'entrepôt de données par ajout d'attributs spécifiques et d'attributs calculés ;
- 4 - Définir les parties historisées de l'entrepôt de données.

Les sections suivantes détaillent chaque étape en présentant l'outil d'aide à l'administrateur ainsi que les concepts associés.

## 4.1 La source globale

L'entrepôt de données est construit à partir d'un schéma source. La source de données repose sur un modèle conceptuel orienté objet qui intègre les concepts de classe et de lien (association, composition, d'héritage). Le schéma source est représenté sous la forme d'un diagramme de classes UML [*UML97*].

Pour illustrer nos propos, nous utilisons un exemple d'application de l'Assurance Maladie en France qui dispose de bases de production volumineuses.

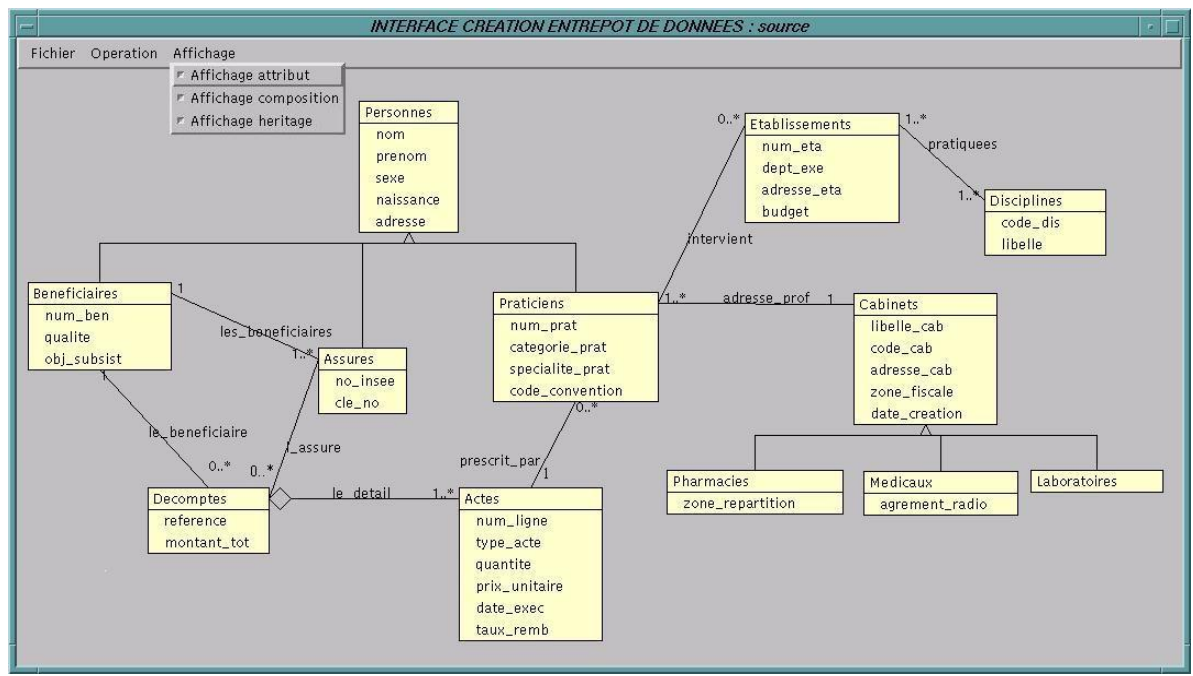

**Figure 3.** *Schéma UML d'une source globale de l'Assurance Maladie.*

Nous proposons une visualisation de la source au travers d'une interface graphique permettant une vision directe du schéma source et une perception intuitive de la sémantique des objets stockés. Afin d'améliorer cette perception (notamment lorsque le graphe est très complexe), différents niveaux d'abstraction sont proposés pour l'affichage du graphe : affichage simple avec le nom des classes et les liens reliant les classes, visualisation des attributs des classes pour affiner la perception du schéma (*cf.* figure 3), simplification du graphe en masquant les liens d'héritage ou/et les liens de composition.

## 4.2 Projection et sélection des classes sources

Nous proposons de définir un entrepôt de données comme une vue portant sur la source globale. La vue est *matérialisée* [*GUPT95*] : les données sont extraites de la source et stockées physiquement dans l'entrepôt. Ainsi les données de l'entrepôt sont disponibles pour une interrogation directe ou pour l'analyse via des outils OLAP (On-line Analytical Processing). Nous adoptons l'approche des vues matérialisées afin de permettre à l'entrepôt de conserver les évolutions des données sources (*cf.* section 4.4).

---

Définition : Une **vue** V est définie comme un schéma de classes dérivées. Le schéma $S^{ED}$ de la vue définissant l'entrepôt de données est représenté par le graphe ($C^{ED}$ ; $L^{ED}$) où

- $C^{ED} = \{C^1, C^2,..., C^n\}$ est un ensemble fini de nœuds représentant les classes dérivées ;
- $L^{ED} = \{L^1, L^2,...L^m\}$ est un ensemble fini d'arcs représentant les liens sémantiques entre classes.

Une **classe dérivée** $C^i$ est définie par ( $Nom^{Ci}$ ; $P^{Ci}$ ; $R^{Ci}$ ; $Ext^{Ci}$ ) où

- $Nom^{Ci}$ est le nom unique de la classe ;
- $P^{Ci}$ est l'ensemble des propriétés qui caractérisent la structure et le comportement des instances de la classe ; $P^{Ci} = A^{Ci} \cup M^{Ci}$ avec $A^{Ci}$ un ensemble d'attributs et $M^{Ci}$ un ensemble d'opérations.
- $R^{Ci}$ est la requête associée à la classe permettant de calculer un ensemble de valeurs ;
- $Ext^{Ci}$ est l'ensemble des instances de la classe obtenues à partir du résultat de la requête.

Un **lien** $L^i$ est défini par ( $CI^{Li}$ ; $CT^{Li}$ ; $\tau^{Li}$ ) où

- $CI^{Li}$ est l'extrémité initiale ;
- $CT^{Li}$ est l'extrémité terminale ;
- $\tau^{Li}$ est le type sémantique (association, composition, héritage).

---

Le schéma des classes dérivées définissant la vue est représenté sous la forme d'un diagramme de classes conforme au formalisme UML [*UML97*].

L'entrepôt est construit par projection (dérivation) d'une partie du graphe source ; l'administrateur projette les classes de la source qu'il souhaite placer dans l'entrepôt. Lors de la projection d'une classe, la classe elle-même, les attributs et les méthodes constituant la classe, ainsi que les liens reliant la classe avec des classes déjà projetées sont projetés.

L'outil garantit la *fermeture du type de la vue* : tous les types nécessaires à la définition de la vue sont présents dans l'entrepôt. Cette propriété de fermeture est obtenue grâce aux principes de la projection automatique des super classes d'une classe projetée ainsi que de ses classes composantes. Par exemple, la projection de la classe "Praticiens" induit la projection de la super classe "Personnes" dont hérite "Praticiens" (*cf.* figure 4).

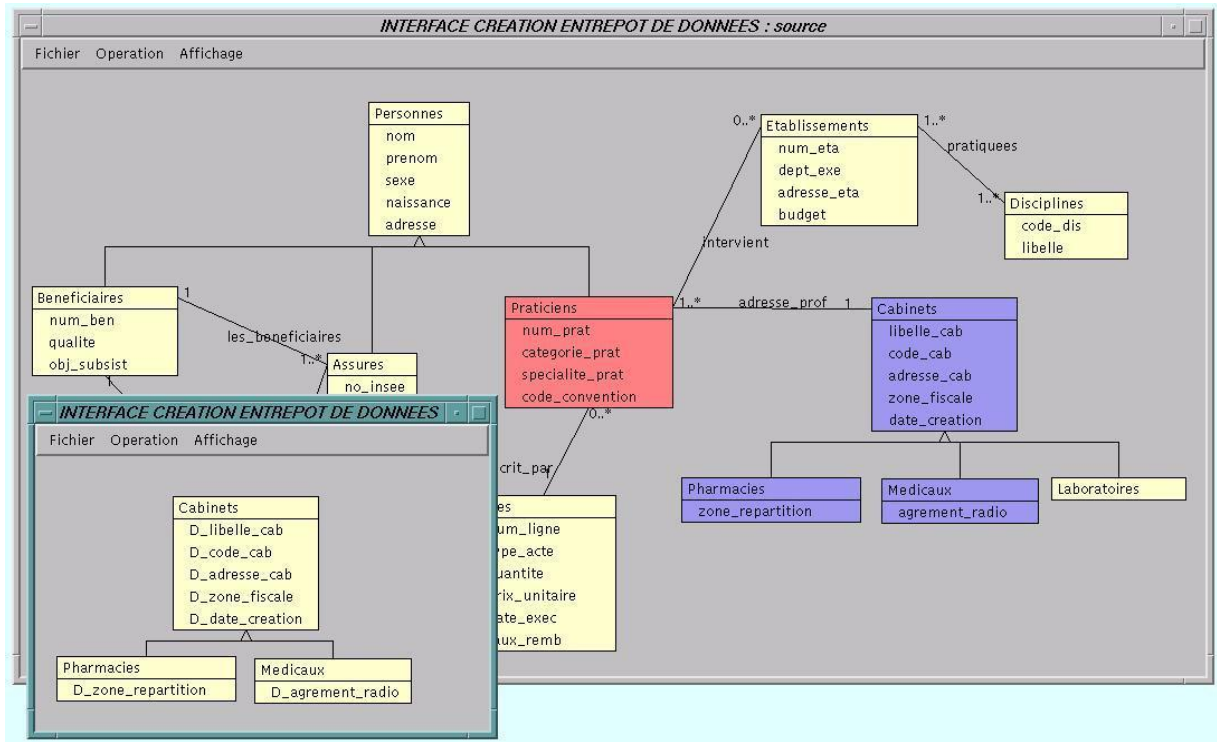

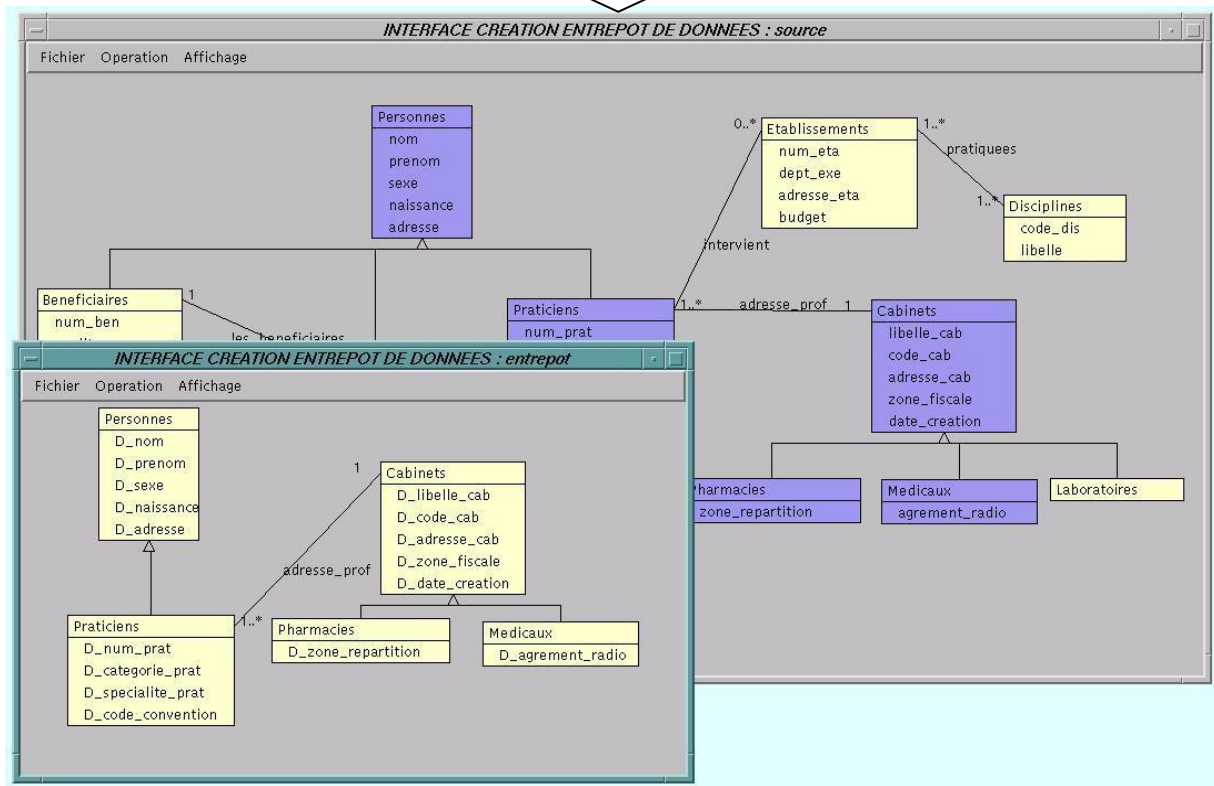

**Figure 4.** *Exemple de projection d'une classe.*

Nous proposons l'opération de *sélection* au niveau des classes source. L'administrateur exprime des sélections afin d'indiquer au système quels objets sources sont dérivés dans l'entrepôt.

Plusieurs opérations permettent de restructurer le schéma de l'entrepôt. Les opérations disponibles sont renommer, supprimer, éclater, regrouper… L'opération regrouper permet de grouper des attributs en un attribut structuré de type n-uplet. L'opération éclater permet d'effectuer l'opération inverse en déstructurant un attribut de type n-uplet.

### 4.3 Enrichissement de l'entrepôt

L'objectif de cette étape est de compléter les données de l'entrepôt. Pour cela, notre modèle distingue trois catégories d'attributs. Certains attributs (attributs dérivés et attributs calculés) sont extraits de la source, et d'autres enrichissent l'entrepôt (attributs spécifiques). Les *attributs dérivés* représentent dans l'entrepôt les attributs sources. Les *attributs calculés* sont obtenus par une combinaison (ou calcul) d'attributs sources. Le calcul est exprimé à l'aide d'une fonction de calcul associée à l'attribut. Les *attributs spécifiques* ne sont pas liés à la source. Ils sont valorisés directement par les utilisateurs ; cela permet de compléter et d'adapter l'entrepôt car la source ne contient pas toujours toute l'information nécessaire aux utilisateurs de l'entrepôt [*CHAU97*].

Nous étendons le formalisme UML en distinguant chaque attribut par l'ajout une préposition : 'D_' pour les attributs dérivés, 'C_' pour les attributs calculés et 'S_' pour les attributs spécifiques.

Par exemple, l'administrateur peut ajouter les attributs spécifiques "S_poids" et "S_taille" à la classe "Personnes" qui permettra aux médecins, utilisateurs de l'entrepôt, de compléter les informations contenues dans cet entrepôt.

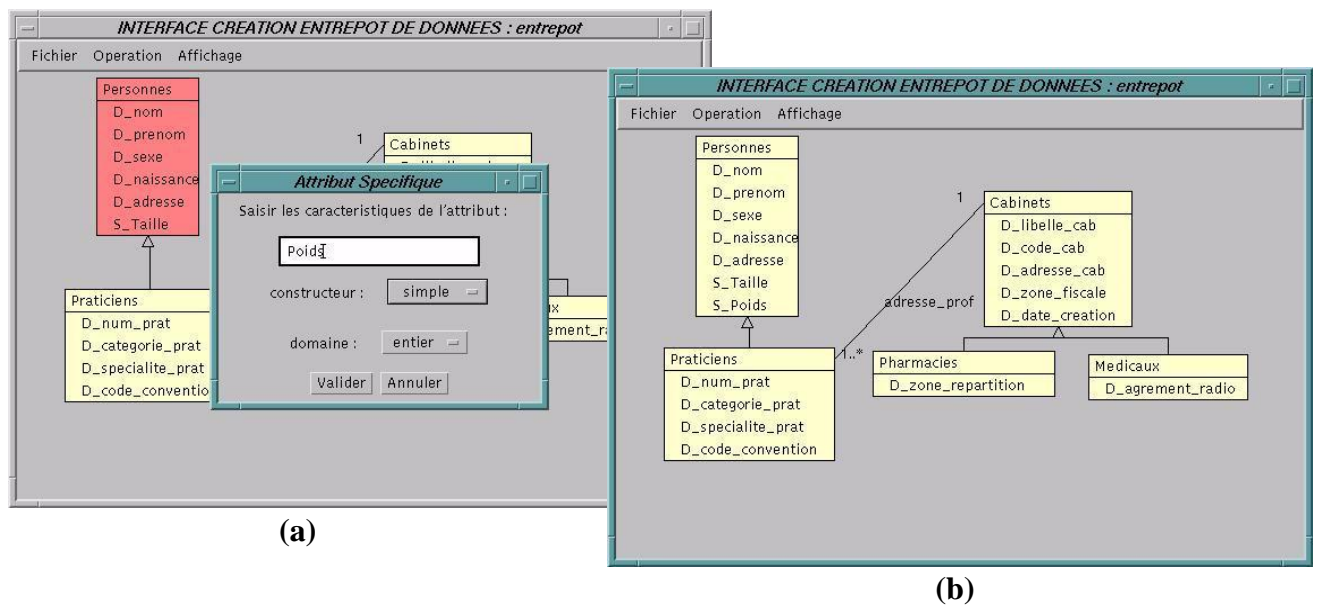

**(a)**

**(b)**

**Figure 5.**   *Création de l'attribut spécifique "P_poids"*.

### 4.4 Historisation des données de l'entrepôt

Une caractéristique importante des entrepôts de données est la possibilité de conserver les évolutions des données. La gestion de ces évolutions nécessite une modélisation du temps dans l'entrepôt. Nous adoptons un modèle temporel proche de [*WANG97*] qui approche le temps par une suite de grains consécutifs disjoints, appelés unités temporelles.

Au cours du temps, les entités du monde réel évoluent. Cette évolution se traduit par des changements de caractéristiques des entités. Dans les bases de données classiques, les entités sont représentées par des objets. Toute évolution des caractéristiques des entités est répercutée dans la BD par la modification de la valeur ou du schéma de l'objet, tandis que les états précédents sont perdus.

Notre modèle permet de gérer des données temporelles à trois niveaux différents, permettant ainsi une grande souplesse dans la construction de l'entrepôt et dans le choix d'historisation des données. Les choix des niveaux temporels sont très importants ; en effet, ces choix détermineront le contenu de l'entrepôt de données et donc détermineront les possibilités d'analyse par rapport au temps. Par ailleurs, les évolutions d'un objet sont décelées uniquement lors de la mise à jour de l'entrepôt (au cours de la phase d'extraction des données sources). La périodicité d'extraction conditionne le contenu de l'entrepôt : une périodicité trop grande par rapport aux évolutions de la source entraîne une perte d'information dans l'entrepôt.

### 4.4.1 Les attributs historisés

Nous proposons un premier niveau d'historisation : celui des attributs. Un attribut historisé est défini par une liste de couples (*val* ; *DateExtrac*) où *val* désigne la valeur prise par l'attribut et *DateExtrac* désigne la date d'extraction de la valeur.

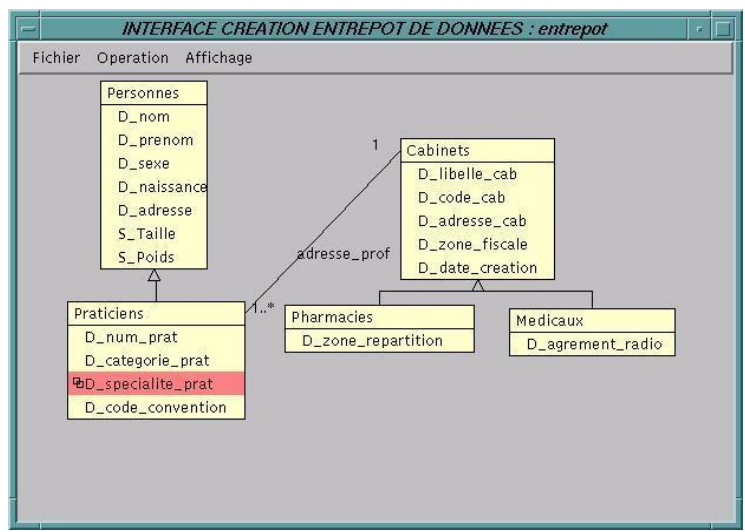

**Figure 6.** *Historisation d'un attribut.*

Par exemple, l'administrateur peut historiser l'attribut "specialite_prat" de la classe "Praticiens" afin de suivre les changements de spécialité des praticiens au cours de leur carrière.

### 4.4.2 Les classes historisées

Dans le cadre de l'historisation des données dans les entrepôts, les évolutions d'une entité peuvent être conservées au niveau de l'objet ; à chaque évolution d'au moins un attribut, un nouvel état est décrit pour l'objet considéré. Un objet dit *générique* contient l'état initial, c'est à dire l'état de l'objet source lors de la première extraction. A chaque extraction, un nouvel état est ajouté dans l'objet générique lorsque l'entité a évolué.

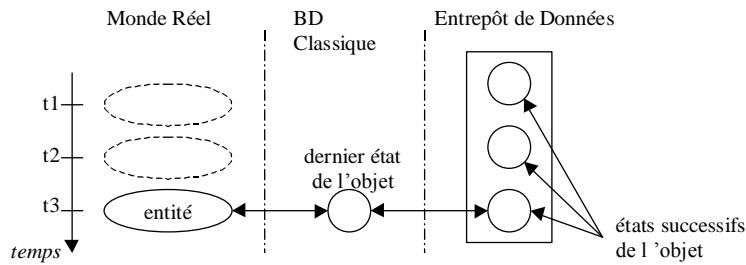

**Figure 7.** *Historisation d'une entité dans un entrepôt.*

Une entité du monde réel est donc décrite par un *objet générique* au niveau de l'entrepôt. Cet objet générique est formé d'un ensemble d'états ; chacun d'eux représentant une valeur de l'entité à un instant donné. L'extension d'une classe historisée est un ensemble d'objets génériques tel que $Ext^{Ci} = \{O^{Ci}_1, O^{Ci}_2,\ldots, O^{Ci}_o\}$.

---

<u>Définition</u> : Un objet générique $OC^i_j$ est défini par $(Id^{Ci}_j, E^{Ci}_j, OS^{Ci}_j)$ où

- $Id^{Ci}_j$ est un identifiant d'objets;
- $E^{Ci}_j = \{(Id^j_1, V^j_1, T^j_1), (Id^j_2, V^j_2, T^j_2),\ldots, (Id^j_x, V^j_x, T^j_x)\}$ est l'ensemble des états de l'objet avec $Id^j_k$ comme identifiant, $V^j_k$ comme valeur et $T^j_k$ comme intervalle de temps.
- $OS^{Ci}_j$ est l'ensemble des objets sources dont dérive l'objet.

---

Par exemple, l'administrateur peut historiser la classe "Beneficiaires" afin de suivre les évolutions de l'ensemble des attributs de la classe.

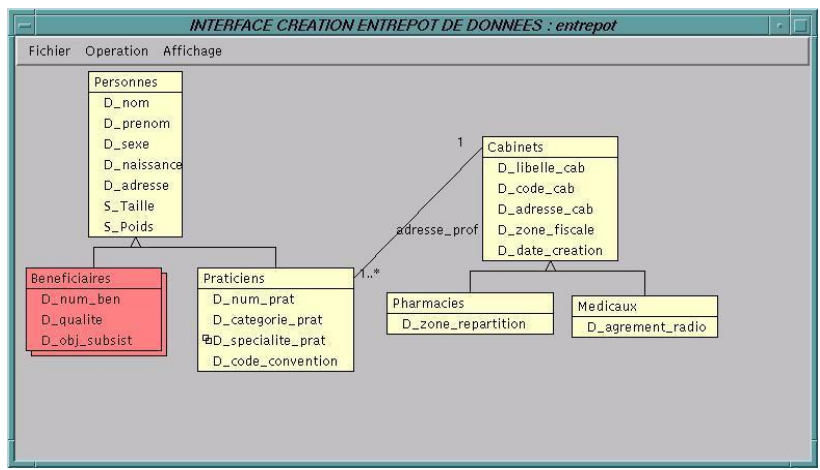

**Figure 8.** *Historisation d'une classe.*

Notons que seulement les évolutions de valeurs des attributs est conservée. Les liens ne sont pas modélisés par des attributs mais sont un concept à part entière du modèle. Les liens relient des objets génériques. Leurs évolutions ne sont conservées que par stockage des évolutions des objets génériques liés (extrémités des liens).

### 4.4.3 Les environnements

Nous proposons un niveau d'historisation plus global concernant un sous-ensemble de classes d'objets et de liens ; on utilise le terme d'*environnement*. La notion d'environnement que nous utilisons s'inspire des versions de base de données [*CELL91*], mais appliquée à un ensemble de classes et non pas à la totalité de l'entrepôt. Un environnement est constitué d'un sous-ensemble des nœuds de $S^{ED}$ (schéma de l'entrepôt) et de certains arcs dont les extrémités initiales et terminales sont des nœuds du sous-schéma.

> <u>Définition</u> : Un environnement $Env^i$ est défini par ($Nom^{Envi}$ , ($C^{Envi}$ ; $L^{Envi}$) ) où
> - $Nom^{Envi}$ est le nom de l'environnement,
> - $C^{Envi} \subseteq C^{ED}$,
> - $L^{Envi} \subseteq L^{ED}\ |\ \forall\ L^{Envi}_k \in L^{Envi},\ CI^{LEnvi} \in C^{Envi} \wedge CT^{LEnvi} \in C^{Envi}$.

Dans notre outil, pour construire un environnement, l'administrateur sélectionne un ensemble de classes et de liens, directement sur le graphe de l'entrepôt avec la souris, puis il définit l'environnement en spécifiant son nom.

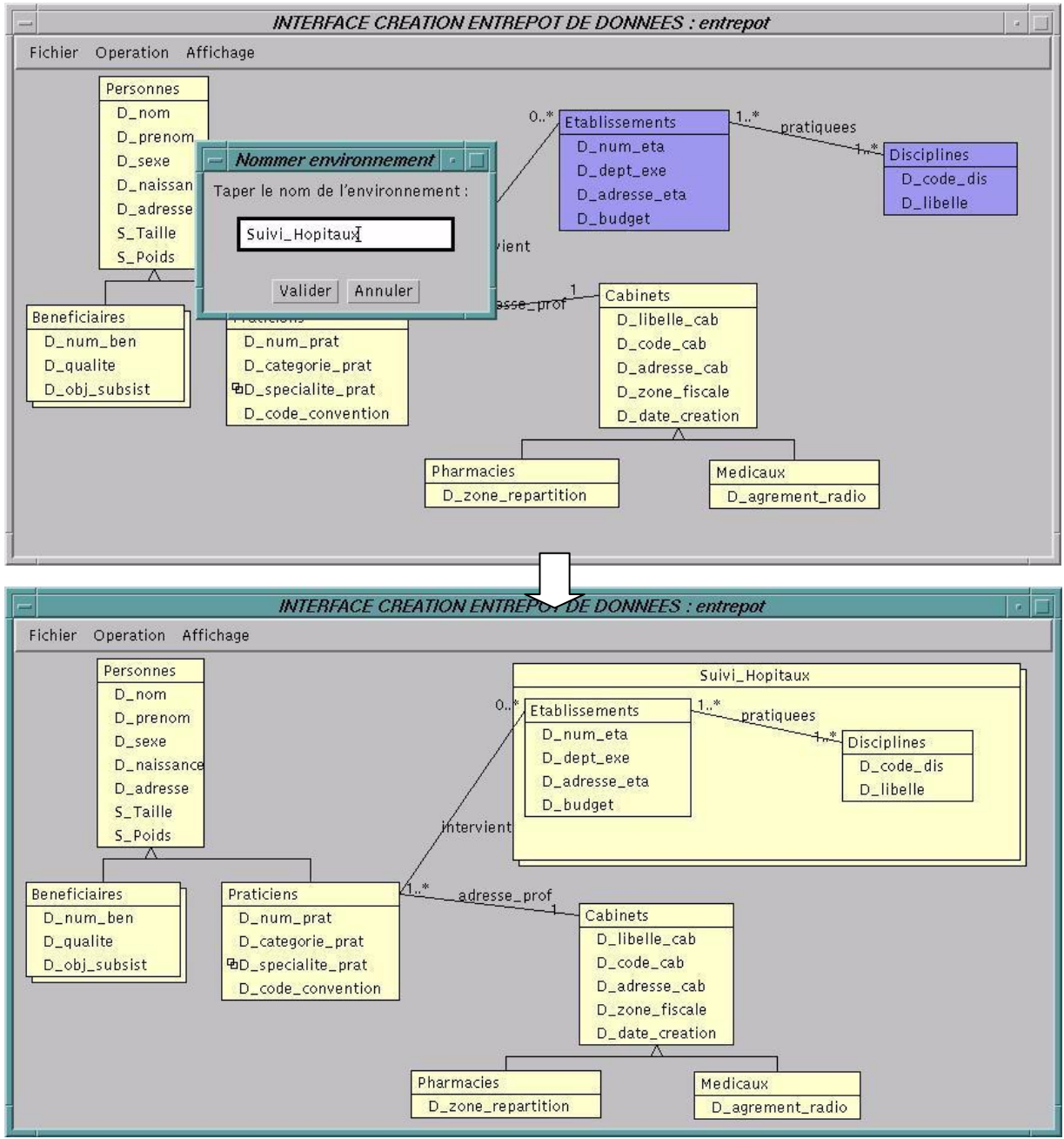

**Figure 9.** *Création d'un environnement.*

L'objectif d'un environnement est de permettre la conservation des évolutions des liens en garantissant la conservation des évolutions des classes d'objets extrémitées des liens. En effet, pour conserver l'évolution d'un lien, il faut conserver les évolutions de ses extrémités.

Plusieurs environnements peuvent être définis, mais leurs ensembles respectifs de nœuds doivent être disjoints : $\forall$ Env$^1$, Env$^2$, Env$^1 \cap$Env$^2=\varnothing$.

# 5 Construction graphique des magasins de données

Le processus OLAP défini par [*CODD93*] consiste à analyser des données suivant différentes dimensions. On peut par exemple analyser les prestations de la sécurité sociale suivant les dimensions "Temps", "Praticiens" et "Bénéficiaires". Une analyse se concentre sur un *sujet* précis. Pour cette raison, les données doivent être organisées de manière particulière. Cette organisation doit mettre en évidence le *sujet analysé* et les *dimensions* de l'analyse. De plus, les temps de réponses doivent être optimisés lors de l'interrogation [*CODD93*]. Ces objectifs sont difficiles à atteindre dans le cadre de l'entrepôt de données où les informations sont exhaustives, détaillées et très normalisées. Ces caractéristiques entraînent une compréhension difficile des informations ainsi que des performances moindres (en particulier à cause de nombreuses opérations de jointures entre classes).

Un magasin contient un sous-ensemble des données de l'entrepôt traitant d'un métier particulier de l'entreprise [*INMO96*]. Il est donc en mesure d'accueillir les données d'un sujet à analyser à condition que celles-ci soient organisées de manière adéquate.

Nous proposons dans cette section une interface graphique d'aide à la construction de magasins de données. Cette interface est basée sur un *modèle conceptuel* et une *démarche*. A l'aide de l'interface, l'administrateur peut *réorganiser* les données de l'entrepôt pour une analyse optimale dans le magasin. Le modèle conceptuel proposé est orienté objet et il étend le principe du schéma en étoile [*SHOS97*].

## 5.1 Détermination des faits à analyser

Le processus OLAP consiste à analyser des mesures d'activité (par exemple la quantité d'actes). Ces mesures sont analysées suivant différents niveaux de détail. On peut ainsi analyser la quantité d'actes pour un mois, un trimestre ou une année avec un niveau de détail décroissant. On passe à un niveau de détail inférieur en regroupant les mesures. Chaque groupe de mesures est ensuite résumé à l'aide de fonctions d'agrégation telles que la somme ou la moyenne. On peut par exemple étudier les quantités d'actes par trimestre en faisant la somme des quantités d'actes par mois.

### 5.1.1 La classe de fait

Les mesures d'activité sont des attributs de la *classe de fait*. Cette classe est l'élément central du magasin de données. Elle se définit comme un ensemble d'attributs $C_f = \{A_1, \ldots, A_m\}$ de type simple. On ne fait pas intervenir les types complexes (utilisant les constructeurs "ensemble" et "liste") pour lesquels des opérations d'agrégation sont difficilement applicables.

### 5.1.2 Projection de la classe de fait à l'aide de l'outil graphique

Dans notre démarche, la classe de fait est *projetée* dans le magasin à partir d'une classe de l'entrepôt représentative du sujet à analyser. L'entrepôt de données peut contenir plusieurs classes représentatives concernant des sujets différents. Pour un sujet donné, on considère une classe représentative unique. Les informations de la classe représentative sont mises à jour fréquemment dans l'entrepôt (en particulier par des opérations d'ajouts) contrairement à des informations plus statiques (celles de la classe "Praticiens" par exemple). Cette classe peut donc être repérée de manière automatique par le système dans l'entrepôt.

A l'aide de l'interface graphique, l'administrateur peut visualiser les classes représentatives candidates présentes dans l'entrepôt de données. Lorsqu'une classe représentative est sélectionnée, elle peut être projetée (avec les attributs qui la composent) dans le magasin de données. Sur l'exemple de la figure 10, la classe représentative de l'analyse des prestations est la classe "Actes". Cette classe est projetée dans le magasin de données en tant que classe de fait "Prestations". Dans le magasin de données, la classe de fait est identifiée par le symbole " 🗇 ". La classe représentative à l'origine d'une classe de fait est identifiée par le même symbole dans l'entrepôt de données.

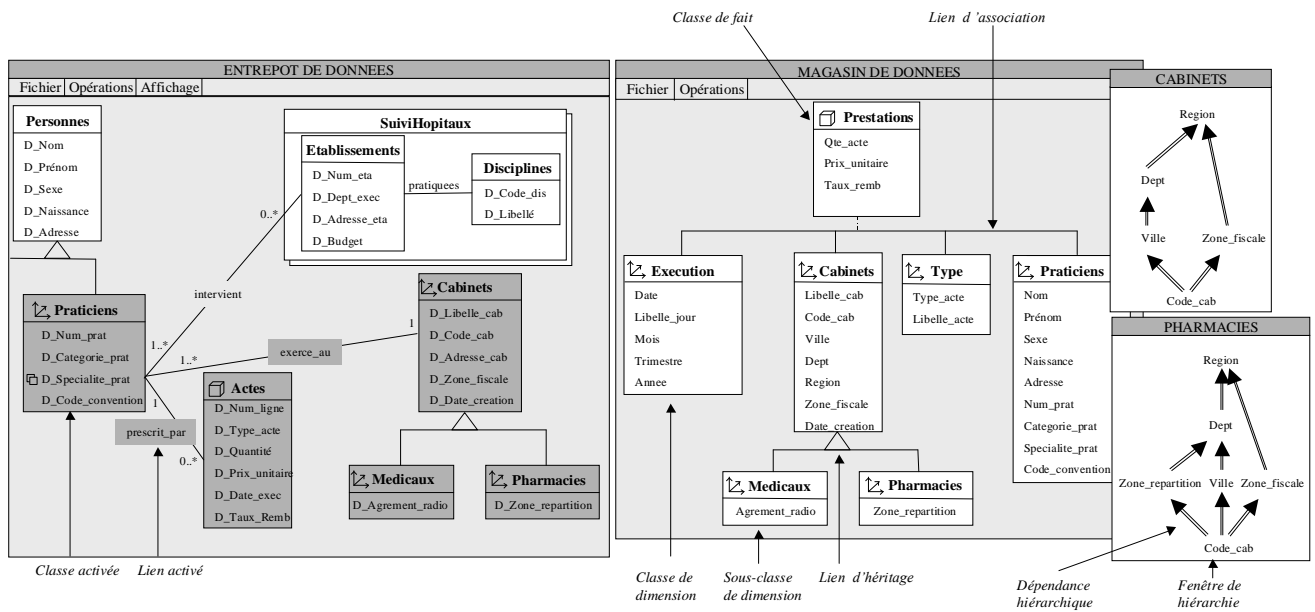

**Figure 10.** *Fenêtres principales de l'outil graphique de conception des magasins de données*

## 5.2 Détermination des dimensions de l'analyse

Les attributs de la classe de fait sont analysés selon certains *paramètres* appartenant aux dimensions. Par exemple, l'attribut "Qté_acte" peut être analysé suivant le mois d'exécution de l'acte (paramètre de la dimension "Execution") ou le type de l'acte (paramètre de la dimension "Type"). Cet exemple est illustré par la figure 10. Les paramètres permettent de déterminer un niveau de détail. Par exemple, une analyse de la quantité d'actes par mois est plus détaillée qu'une analyse par trimestre.

### 5.2.1 Les classes de dimension

Une *classe de dimension* se définit comme un ensemble d'attributs Cd = {$A_1$, …, $A_d$}. Ces attributs peuvent être de type simple (chaîne de caractère, numérique) ou de type complexe (ensemble, liste). Les types complexes permettent d'exprimer des requêtes d'analyse difficilement exprimables dans un contexte relationnel ou multidimensionnel classique (avec des quantificateurs "Au moins" et "Tous"). La classe de fait est reliée à une classe de dimension par un lien d'association de cardinalité (1,1). On peut ainsi considérer la classe de fait comme une *classe d'association* [*UML97*] reliant les classes de dimension (*cf.* figure 10).

Sur chaque classe de dimension est définie une hiérarchie $H_d$ des attributs. Les attributs d'une hiérarchie sont des paramètres de l'analyse. Ils sont liés par des dépendances hiérarchiques. Sur l'exemple de la figure 10, l'attribut "Ville" de la classe de dimension "Cabinets" dépend hiérarchiquement de l'attribut "Département".

---
Définition : Une **dépendance hiérarchique** entre deux attributs $A_i$ et $A_j$ (notée $A_i \Rightarrow A_j$) implique :

- une dépendance fonctionnelle entre $A_i$ et $A_j$ (notée $A_i \rightarrow A_j$),
- l'absence d'une dépendance fonctionnelle inverse $A_j \rightarrow A_i$.

---

On peut ainsi interpréter une dépendance hiérarchique entre $A_i$ et $A_j$ comme : à une valeur de $A_i$ est associée exactement une valeur de $A_j$ et à une valeur de $A_j$ est associée une ou plusieurs valeurs de $A_i$. Ainsi, pour chaque attribut $A_i \in H_d$, il existe un attribut $A_j$ tel que $A_i \Rightarrow A_j$ ou $A_j \Rightarrow A_i$.

---
Définition : Une **hiérarchie de dimension** est représentée par un graphe acyclique $G_h = (N_h, L_h)$ ou $N_h$ est un ensemble de nœuds et $L_h$ un ensemble de liens entre ces nœuds. Chaque nœud est associé à un attribut de la hiérarchie. Un lien correspond à une dépendance hiérarchique entre deux attributs.

---

Les classes de dimension peuvent être reliées par un lien d'héritage. Un lien d'héritage permet de spécialiser une classe de dimension. La sous-classe obtenue possède tous les attributs de la super-classe, plus des attributs supplémentaires. On peut ainsi spécialiser une hiérarchie. (chaque sous-classe peut posséder sa propre hiérarchie). Sur l'exemple de la figure 10, la classe de dimension "Pharmacie" est une sous-classe de la classe de dimension "Cabinets". Cette sous-classe possède sa propre hiérarchie qui spécialise la hiérarchie de la super-classe "Cabinets".

---
Définition : Le principe de **dépendance entre classes** est utilisé dans l'entrepôt de données pour déterminer les classes de dimension d'une analyse. Une classe $C_2$ est dépendante d'une classe $C_1$ ($C_1 \Rightarrow C_2$) dans l'un des cas suivants :

- $C_1$ est reliée à $C_2$ par un lien d'association de cardinalité (x,1),
- $C_1$ est reliée à $C_2$ par un lien d'héritage (dans le sens C1 est une super-classe de C2 ou dans le sens C1 et une sous-classe de C2),
- $C_1$ est reliée à $C_2$ par un lien de composition (dans le sens $C_1$ est une classe composante de la classe composée $C_2$ ou dans le sens $C_2$ est une classe composée de la classe composante $C_1$ avec une cardinalité (x,1)),
- $C_1 = C_2$.

---

Le principe de *transitivité* des dépendances se définit de la manière suivante : $C_1 \Rightarrow C_2$ et $C_2 \Rightarrow C_3$ permet de déduire que $C_1 \Rightarrow C_3$. Sur l'exemple de la figure 10, la classe "Praticiens" est dépendante de la classe "Actes". De même, la classe "Cabinets" est dépendante de la classe "Praticiens". Par transitivité, la classe "Cabinets" est donc dépendante de la classe "Actes".

### 5.2.2 *Projection des classes de dimension à l'aide de l'outil graphique*

Dans notre démarche, on peut déterminer toutes les classes dépendantes de la classe représentative en utilisant le principe de transitivité des dépendances. Les classes dépendantes de la classe représentative sont projetées (avec leurs attributs) dans le magasin en tant que

*classe de dimension*. L'administrateur peut choisir les classes dépendantes à projeter ou laisser le système projeter toutes les classes dépendantes détectées.

On peut également projeter une classe de dimension dans le magasin à partir d'un attribut d'une classe dépendante. Un attribut représentant une date peut ainsi être à l'origine d'une classe de dimension temporelle. De même, un attribut représentant une adresse peut être à l'origine d'une classe de dimension géographique. Sur l'exemple de la figure 10, l'attribut "Date_exec" de la classe "Actes" est à l'origine de la classe de dimension "Execution".

L'administrateur définit ensuite la hiérarchie de chacune des classes de dimension du magasin de données. Cette étape de la démarche peut être assistée par le système qui recherche les dépendances hiérarchiques entre attributs de la classe de dimension. Ces attributs sont ensuite ordonnés pour définir les différents niveaux de détail de la dimension. L'administrateur peut également définir lui-même une hiérarchie si les dépendances sont définies dans un dossier d'analyse ou si elles sont présentes sous forme de métadonnées dans l'entrepôt.

L'interface graphique d'aide à la construction de magasins de données propose automatiquement à l'administrateur les classes dépendantes d'une classe sélectionnée. L'administrateur active une ou plusieurs classes dépendantes en empruntant un lien. Sur l'exemple de la figure 10, depuis la classe "Actes", le concepteur active la classe dépendante "Praticiens" en empruntant le lien "Prescrit_par". Les classes et les liens activés sont grisés sur le graphe représentant le schéma de l'entrepôt. Dans le magasin de données, une classe de dimension est identifiée par le symbole " ↗ ". La classe dépendante à l'origine d'une classe de dimension est identifiée par le même symbole dans l'entrepôt de données.

L'interface d'aide à la construction de magasin propose également une fenêtre de hiérarchie permettant d'organiser les attributs d'une classe de dimension suivant leurs dépendances hiérarchiques. La figure 10 présente la fenêtre de hiérarchie de la classe de dimension "Cabinets".

## 5.3 Opérations sur les classes du magasin de données

Lorsque la classe de fait et les classes de dimension sont projetées dans le magasin de données, l'administrateur se trouve en présence d'un schéma orienté objet en étoile. Les opérations présentées dans ce paragraphe permettent de modifier ce schéma (opération d'ajout ou de suppression de mesures ou de paramètres) ou de sélectionner des objets de classes (opération de sélection) pour limiter les objets importés depuis l'entrepôt de données dans le magasin.

### 5.3.1 *Manipulation d'attributs calculés*

On peut ajouter un attribut calculé à une classe du magasin de données. Cet attribut est une combinaison d'attributs appartenant aux classes de l'entrepôt. Un attribut calculé est associé à une formule de calcul qui se présente sous la forme d'un arbre $G_c=(N_c, L_c)$. $N_c$ est un ensemble de nœuds. Chaque nœud représente un opérateur ou une opérande. Les opérateurs peuvent être de type scalaire (addition, soustraction, multiplication, division) ou agrégation (somme, moyenne, comptage, min, max). Les opérandes sont des attributs des classes de l'entrepôt. $L_c$ est un ensemble de liens entre les nœuds. Ces liens permettent d'associer opérateurs et opérandes.

Dans notre démarche, l'administrateur peut décider de supprimer des attributs de la classe de fait ou des classes de dimension qui n'intéressent pas l'analyse. Il peut également créer des mesures calculées dans la classe de fait. On peut ainsi analyser des informations qui ne sont pas directement présentes dans l'entrepôt. Une mesure calculée est exprimée sous forme d'une requête s'appliquant sur l'entrepôt. Le point de départ de cette requête est la classe représentative à l'origine de la classe de fait. A partir de cette classe, on accède aux classes voisines via les liens. On peut ainsi créer une mesure calculée "Montant remb" dans la classe de fait. Les valeurs de cette mesure sont calculées suivant la formule suivante : ("Actes.Quantité" * "Actes.Prix Unitaire") * "Actes.Taux Remb". Chaque nom d'attribut intervenant dans la formule est précédé du nom de la classe à laquelle il appartient.

Selon un principe similaire, l'administrateur peut créer des *paramètres calculés* dans une classe de dimension. Un paramètre calculé s'exprime également sous forme d'une requête appliquée à l'entrepôt. Le point de départ de cette requête est la *classe dépendante* à l'origine de la classe de dimension. A partir de cette classe, on accède aux classes voisines via les liens. Les paramètres calculés permettent par exemple d'ajouter des paramètres "Mois" et "Année" à partir d'un paramètre "Date" d'une classe de dimension "Temps". De même, à partir d'un paramètre "Adresse" d'une classe de dimension "Cabinets", on peut créer des paramètres calculés "Ville", "Département" et "Région". Sur l'exemple de la figure 11, les paramètres "Libelle_jour", "Mois", "Trimestre" et "Annee" de la classe de dimension "Execution" sont calculés à partir de l'attribut "Date_exec" de la classe "Actes" présente dans l'entrepôt.

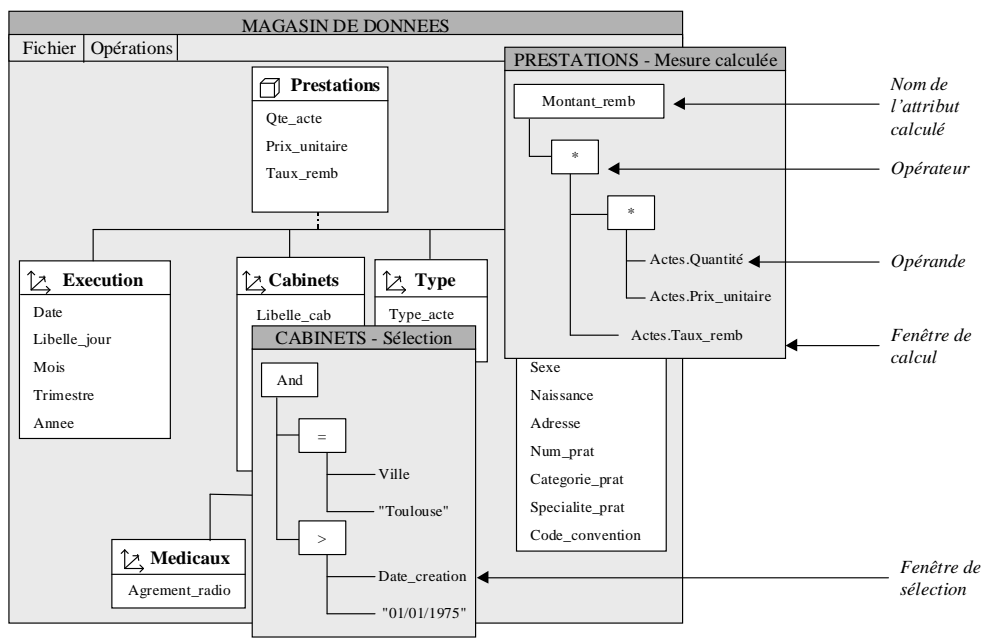

**Figure 11.** *Opérations principales dans un magasin de données*

Pour ajouter des mesures ou des paramètres calculés, notre interface propose une fenêtre de calcul. Cette fenêtre permet à l'administrateur de construire une formule de calcul. Dans la figure 11, l'administrateur ajoute une mesure calculée "Montant_remb" à la classe de fait.

### 5.3.2 *Opération de sélection*

L'opération de sélection permet de construire un sous-ensemble d'objets d'une classe de fait ou d'une classe de dimension. L'administrateur peut par exemple décider de ne conserver dans le magasin que les cabinets de la ville de Toulouse créés après le 1er janvier 1975. Une formule de sélection est associée à une opération de sélection d'objets. Ce type de formule se présente sous la forme d'un arbre. Chaque nœud correspond à un opérateur ou une opérande. Les opérateurs permettent de créer des expressions booléennes (et, ou, non) ou de comparaison (égal, supérieur, inférieur). Les opérandes sont des attributs d'une classe de l'entrepôt.

L'opération de sélection est disponible dans notre outil graphique via une fenêtre de sélection (*cf.* figure 11).

# 6 Conclusion

Dans cet article, nous avons présenté notre contribution en matière de construction d'entrepôts et de magasins de données. En particulier, nous avons proposé :

- une architecture des systèmes décisionnels à quatre niveaux permettant de distinguer clairement les problématiques liées à chaque phase de construction (intégration des sources de données, construction de l'entrepôt de données, construction des magasins de données).
- une démarche et une interface graphique qui assiste l'administrateur pour construire l'entrepôt de données. Notre approche est basée sur un modèle d'entrepôt de données orientées objet intégrant un mécanisme d'historisation des données à trois niveaux (attribut, classe et environnement).
- une démarche et une interface graphique qui assiste l'administrateur dans la phase de réorganisation permettant de construire des magasins de données. Ce processus est basé sur un modèle orienté objet multidimensionnel.

Nous avons développé un prototype au dessus du SGBD O2 permettant de valider nos propositions. Pour cela, nous avons défini un méta-schéma permettant la gestion des entrepôts de données dans le SGBD O2 ; tout entrepôt défini graphiquement par une vue est généré à partir des classes du méta-schéma grâce au mécanisme d'héritage. A partir des spécifications graphiques, notre système génère les scripts de création de la structure de l'entrepôt ou du magasin, mais aussi les scripts de chargement initial et de rafraîchissement des données. Nous avons implanté une interface administrateur permettant de définir graphiquement une vue et d'historiser l'entrepôt à différents niveaux (attribut, classe, environnement). Une autre interface permettant de construire des magasins de données est en cours de réalisation. Cette interface permet de réorganiser les données de l'entrepôt sous forme multidimensionnelle.

Nous souhaitons poursuivre nos travaux en proposant des langages graphiques permettant la manipulation et l'interrogation des données au niveau de l'entrepôt et la manipulation et l'interrogation multidimensionnelle des données au niveau des magasins. Nous allons également compléter nos travaux en intégrant des contraintes dans notre modèle d'entrepôt de données. Ces contraintes permettront à l'administrateur de configurer le processus d'historisation des données de l'entrepôt.

# 7 Remerciements